\documentclass[aps,prl,reprint,superscriptaddress]{revtex4-1}
\usepackage{amsmath}
\usepackage{graphicx,dcolumn,bm,natbib,epsfig,verbatim,color,soul}
\usepackage{bm}        
\usepackage{mathrsfs} 
\usepackage{color}
\usepackage{subfigure}  
\usepackage{graphicx}

\def\ta0{\tilde a_0}
\def\ta{\tilde a}
\def\be{\begin{equation}}
\def\ee{\end{equation}}
\def\bc{\begin{columns}}
\def\ec{\end{columns}}
\def\bea{\begin{eqnarray}}
\def\eea{\end{eqnarray}}

\def\3dots{\:\raisebox{-0.5ex}{$\stackrel{\textstyle.}{:}$}\:}

\begin{document}
\title{Rigidity and fracture of fibrous double networks}

\author{Pancy Lwin}
\affiliation{School of Physics and Astronomy, Rochester Institute of Technology, Rochester, NY 14623, USA}
\author{Andrew Sindermann}
\affiliation{School of Physics and Astronomy, Rochester Institute of Technology, Rochester, NY 14623, USA}
\author{Leo Sutter}
\affiliation{School of Physics and Astronomy, Rochester Institute of Technology, Rochester, NY 14623, USA}
\author{Thomas Wyse Jackson}
\affiliation{Department of Physics, Cornell University, Ithaca, NY 14853, USA}
\author{Lawrence Bonassar}
\affiliation{Meinig School of Biomedical Engineering and Sibley School of Mechanical and Aerospace Engineering, Cornell University, NY 14853, USA}
\author{Itai Cohen}
\affiliation{Department of Physics, Cornell University, Ithaca, NY 14853, USA}
\author{Moumita Das}
\affiliation{School of Physics and Astronomy, Rochester Institute of Technology, Rochester, NY 14623, USA}
\date{\today}

\begin{abstract}
Tunable mechanics and fracture resistance are hallmarks of biological tissues and highly desired in engineered materials. To elucidate the underlying mechanisms, we study a rigidly percolating double network (DN) made of a stiff and a flexible network. The DN shows remarkable tunability in mechanical response when the stiff network is just above its rigidity percolation threshold and minimal changes far from this threshold. Further, the DN can be modulated to either be extensible, breaking gradually, or stronger, breaking in a more brittle fashion by varying the flexible network's concentration. 
 \end{abstract}

\maketitle
Composite fiber networks are ubiquitous in biological systems and synthetic materials with tunable and robust mechanical properties.  For example, the cytoskeleton, the scaffolding that gives eukaryotic cells mechanical integrity and shape, is a self-organized composite network of protein filaments, including actin and microtubules \cite{Cytoskeleton}. The distinct rigidity of actin and microtubules enables cells to exhibit complex stress responses and architectures essential for a wide range of functions \cite{Pelletier,Ricketts2018,Ricketts2019,Farhadi}. As another example, the load-bearing capability of musculoskeletal tissues such as articular cartilage arises from a network-like extracellular matrix made of collagen fibers and proteoglycans \cite{Mow:1992,Fung:1993,ODriscoll:1998,Lai:2010}. Finally, several synthetic double network hydrogels have recently emerged as extraordinarily robust materials with considerable toughness and fracture resistance compared to conventional single network hydrogels. For instance, the PAMPS-PAAm double network hydrogel, which consists of interacting networks of poly(2-acrylamide-2-methyl-propane sulfonic acid) and polyacrylamide, has a tearing energy~$\sim 4400J/m^2$, which is several hundred to a thousand times that of single network PAAm and PAMPS hydrogels\cite{Gong2012,Gong2015}. The exceptional mechanical response of these double network systems derives from the synergistic interplay between two networks with very different single-filament and collective properties.

 \begin{figure}
\includegraphics[width=\columnwidth]{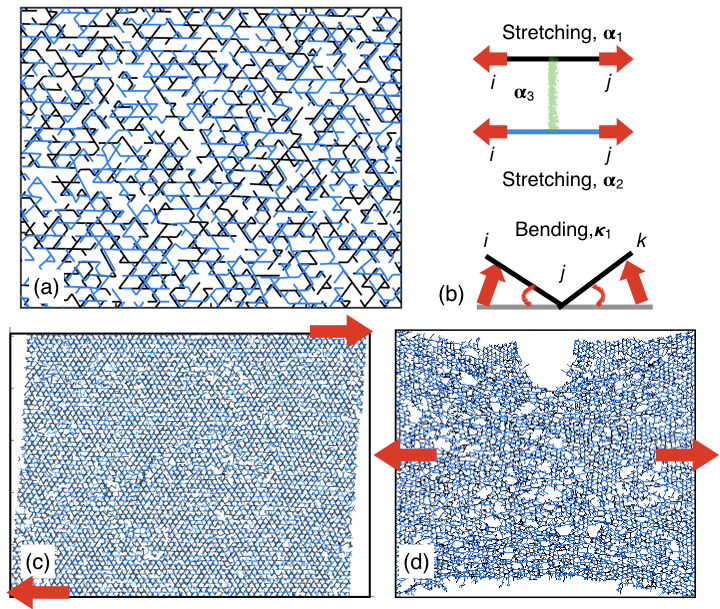}
\caption{Figure (a) represents a schematic of a zoomed-in portion of the DN, and (b)  the different contributions to its deformation energy. The black and blue fibers belong to the the stiff and flexible networks respectively. Figures (c) and (d) show representative DNs (with $p_1=0.62, p_2=0.6$) for our studies of shear response and crack propagation respectively.}
\label{schematic}
\end{figure}

The rigidity of stiff networks made of a single type of fiber or biopolymer, henceforth called single networks (SN), have been vigorously investigated in the past two decades, uncovering mechanical phase transitions, distinct mechanical regimes, and novel non-linear mechanical properties \cite{Broedersz:2011,Head:2003,Wilhelm:2003,gardel2004scaling,Missel:2010,Heussinger:2006,Das:2007,kang2009nonlinear,Sheinman:2012,Picu:2011}. More recently, studies of the fracture mechanics of such networks have demonstrated that low network connectivity and system-wide distribution of damage can provide protective mechanisms against failure \cite{Zhang2019,Burla2020}. The mechanics and fracture of composite networks, on the other hand, are only beginning to be explored, and theoretical studies to date have focused on composites made of rod-like inclusions in an SN \cite{Das:2008, Das:2010}, bidispersity in single networks \cite{Burla2019}, and continuum models of double network hydrogels \cite{Gong2012}. The mechanical structure-function properties of double networks (DN) are less well understood, and there remain many open questions as to the mechanisms by which DNs achieve such remarkable mechanical performance. In particular, it is unknown how much the second network can affect the rigidity percolation threshold for the combined DN system, an important parameter for setting the stiffness. Nor is it known to what degree the second network can tune the strain necessary for network failure (extensibility), the maximum stress reached (strength), and the toughness under extension, all of which are important for determining the workable range of strains and stresses over which the system maintains its integrity. Addressing these questions will help guide the rational design of biomimetic soft materials with tunable mechanics and provide insights into the rigidity and fracture-resistance of load-bearing tissues such as cartilage \cite{Trivedia:2008,Simms}. 
 
Here, we address these questions by combining two structure-function frameworks, (i) a double network (DN) made of two interacting disordered networks with very distinct mechanics and (ii) rigidity percolation theory to construct a Rigidly Percolating Double Network model.  Rigidity percolation theory models a biopolymer network as a disordered network of fibers consisting of flexible, sparsely connected regions and stiff, densely connected regions \cite{silverberg2014structure,FengThorpe,Das:2007,Das:2012,Broedersz:2011}.  When the network consists primarily of sparsely connected regions, it does not offer any resistance to shear deformations and has zero shear modulus. In contrast, when densely connected regions span the network, the network has a finite shear modulus. The system undergoes a mechanical phase transition from non-rigid to rigid at a certain fiber density known as the rigidity percolation threshold. 

The rigidly percolating double network model is made of a stiff network interacting with a flexible network (Fig.\ref{schematic}(a)). We study the shear response and crack propagation in this DN and show that the interplay of the mechanically distinct networks facilitates tunable mechanics and enhanced fracture resistance of the DN. Each of the two networks in the DN is modeled as a disordered kagome network and is constructed following the procedure described in \cite{silverberg2014structure}. The bonds in the two networks are randomly removed according to two different probabilities, $1-p_1$ for the stiff network, and $1-p_2$ for the flexible network, where $0 <p_1,p_2 <1$ are the bond occupation probabilities. The energy cost of deforming this double network is given by: 
\begin{eqnarray} 
E_{1}&=&\frac{\alpha_{1}}{2}\sum_{<ij>}^{}p_{1,ij} (\textbf{r}_{ij} - \textbf{r}_{ij0})^{2}  \nonumber \\
         &+& \frac{\kappa_{1}}{2}  \sum_{<\widehat{ijk}=\pi>}  p_{1,ij} \; p_{1,jk} \;\; \Delta \theta_{ijk}^{2}  \nonumber \\
E_{2}&=&\frac{\alpha_{2}}{2}\sum_{<ij>}^{}p_{2,ij} ( \textbf{s}_{ij} - \textbf{s}_{ij0})^{2} \nonumber \\
E_{3}&=&\frac{\alpha_{3}}{2}\sum p_{1,ij} \; p_{2,ij}  (\textbf{x}_{1}-\textbf{x}_{2})^{2}, 
\label{sim1}
\end{eqnarray}
where $E_1$ is the deformation energy of the stiff network, $E_2$ is the deformation energy of the flexible network, and $E_3$ is the deformation energy of the bonds connecting the two networks. In $E_1$, the first term corresponds to the energy cost of fiber stretching, and the second term to fiber bending \cite{silverberg2014structure}. In $E_2$, we have similar contribution for fiber stretching, but there is no energy cost of fiber bending. 
The stretching modulus of the fibers in the stiff and flexible networks are $\alpha_1$ and $\alpha_2$ respectively where $\alpha_1 > \alpha_2$,  the bending modulus of the fibers in the stiff network is $\kappa_1$.  The networks interact via Hookean springs with spring constant $\alpha_3$ which connect the midpoints of bonds ($\textbf{x}_{1}$,$\textbf{x}_{2}$), and are only present when corresponding bonds are present in both networks. The indices $i,j,k$ refer to sites (nodes) in each lattice based network, such that $p_{ij}$ is 1 when a bond between those lattice sites is present, 0 if a bond is not present. The  quantities $\textbf{r}_{ij}$ and $\textbf{s}_{ij}$ refer to the vector lengths between lattice sites $i$ and $j$ for the deformed stiff and flexible networks respectively, while $\textbf{r}_{ij0}$ and $\textbf{s}_{ij0}$ are the corresponding quantities for the initial undeformed networks. The angles $\Delta \theta_{ijk}$ correspond to the change in the angles between initially collinear bond pairs $ij$ and $jk$ for the deformed and undeformed network respectively.  See Figure \ref{schematic} (b) for illustration of the properties of the bonds in the networks. Unless otherwise noted, we have used the following biologically relevant parameters in the results presented: $\alpha_2/\alpha_1=0.1$, $\kappa/\alpha_1=.004$ \cite{silverberg2014structure}, and $\alpha_3=\alpha_1 + \alpha_2$. \footnote{The value of $\alpha_3$ was chosen to be the effective spring constant of two springs $\alpha_1$ and $\alpha_2$ in parallel. Simulations with a much small value of $\alpha_3/\alpha_1=0.01$ yielded qualitatively similar results and the percolation thresholds were unchanged.}
 
  \begin{figure}
\includegraphics[width=0.9\columnwidth]{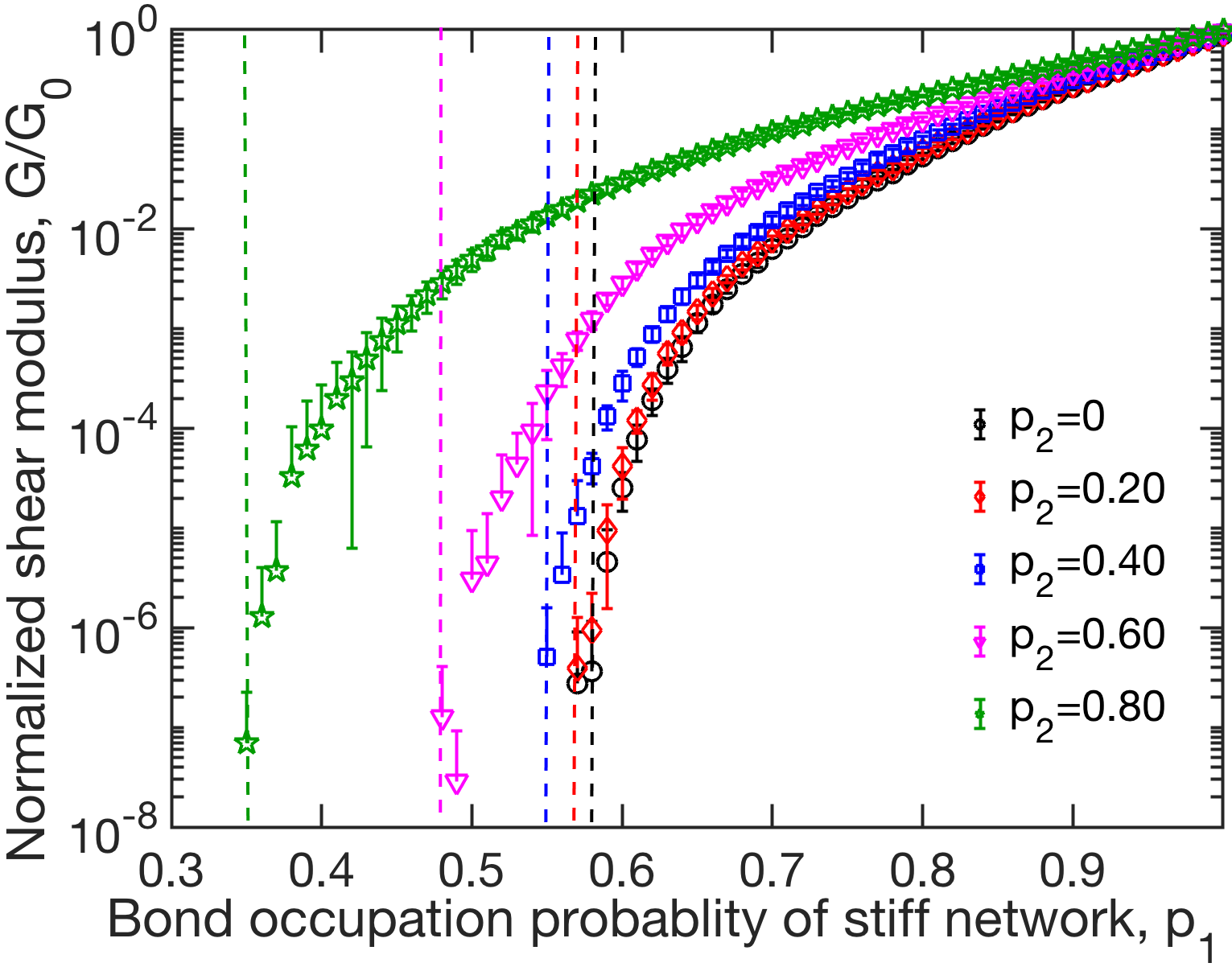}
\caption{The normalized shear modulus  ($G/G_0$)shown as a function of $p_1$ for an SN (black circles) and four DNs (remaining data). The values of $p_2$ are shown in the legend. The dashed lines provide guide to the eye for the rigidity percolation transitions. The data is averaged over five runs and the standard deviations are indicated by errorbars.}
	\label{shearmod}
\end{figure}

For the shear response studies (Fig.~\ref{schematic}(c)), we adopt a protocol where external deformations are applied along the top and bottom boundaries and periodic boundary conditions are used for the left and right sides of the network.  Our simulations of the single network follow the same process, except the deformation energy consists of only $E_1$, since $p_2=0$. To obtain the linear mechanical response, we apply a shear strain of $5\%$ at the boundaries, minimize the deformation energy using a conjugate minimization scheme and calculate the shear modulus \cite{silverberg2014structure}. 

We show the variation of the rigidity percolation threshold of a single network (SN) and four double networks (DN) by plotting the shear modulus versus bond occupation probability of the stiff network $p_1$ (Figure \ref{shearmod}).  The shear moduli $G$ are normalized by their respective values $G_0$ for fully occupied networks. The four DNs correspond to the four different values of the bond occupation probability $p_2=0.2,0.4,0.6,0.8$ of the flexible network. We find that the SN has a percolation threshold $p_{1,c} \sim 0.6$ in agreement with previous results \cite{Lubensky2013},  while the DNs have a lower $p_{1,c}$, which decreases with increasing $p_2$, reaching $p_{1,c} \sim 0.35$ at $p_2=0.8$. This is a noteworthy result, because on their own a single stiff network based on a Kagome lattice has a percolation threshold $\sim 0.6$ \cite{Lubensky2013} and a flexible network based on such a lattice has a percolation threshold $\sim 1$, but when they form a double network, the resulting additional constraints due to their interaction lead to a lower, tunable percolation threshold. These constraints also allow the normalized shear rigidity of the DNs to be larger than that of the SN at the same value of $p_1$, and can be tuned by varying $p_2$. This result illustrates a mechanism for how the onset of rigidity for biological and synthetic double networks can be drastically modulated through very small changes in filament concentration in the secondary network.  

For the crack propagation studies (Fig.~\ref{schematic}(d)), we create a notch 20 times the bond rest length ($\sim1/5$ times the system length) at the center of the top boundary following the protocol in Ref.\cite{Oloyede2003}, and study how the size of the notch increases as we apply larger and larger tensile strains along the left and right boundaries of the network. 

The strains are applied quasi-statically in small increments of $1 \%$ up to $70\%$, and after each application, the total energy is minimized to generate the new equilibrium configuration of the deformed DN. We have made the following assumptions regarding breaking and buckling of fibers:  When a bond is stretched above a certain threshold, it will break, and when it is compressed above a certain threshold, it will buckle. Bonds in the stiff network break at $120 \%$ of their rest length and buckle at $95\%$ of their rest length. Bonds in the flexible network break at $200\%$ of their rest length, but do not buckle.  Broken or buckled bonds will no longer contribute to the deformation energy or rigidity of the network. Fiber breaking is an irreversible process, but buckled fibers in our model can ``unbuckle" when the extra compression is removed. 

\begin{figure}
	\includegraphics[width=1 \columnwidth]{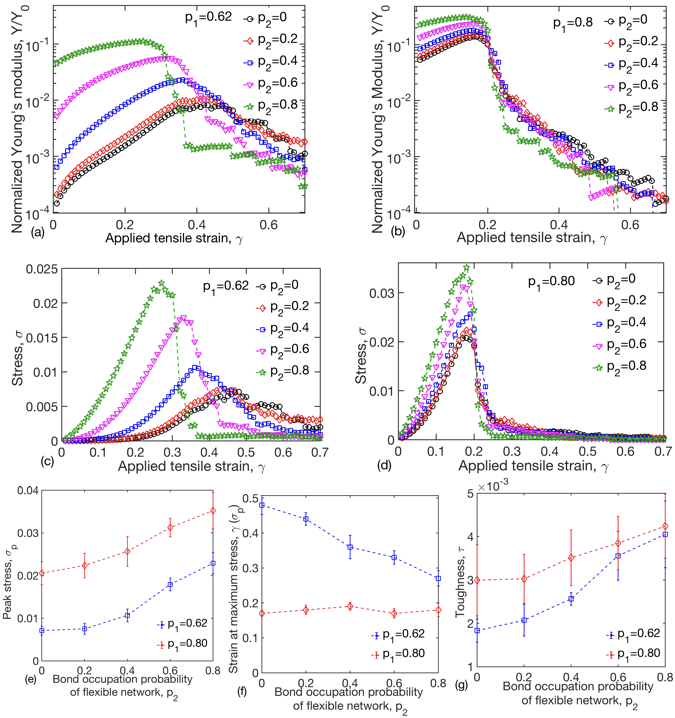}
	\caption{Figures (a-b) show the normalized Young's modulus $Y/Y_0$  and (c-d) show the stresses $\sigma$ developed in the SN (black circles) and  DN (remaining data) as a function of the uniaxial tensile strain $\gamma$ applied at the boundaries. Figure (a) and (c) corresponds to $p_1=0.62$, and figure (b) and (d) to $p_1=0.80$; $p_2$ is as shown in the legend in these figures. Figures (e), (f), and (g) show the peak stress ($\sigma_p$), and the strain at peak stress ($\gamma (\sigma_p)$), and the toughness ($\tau$) as a function for $p_2$ for the data shown in (c) and (d). The stress is expressed in units of $\alpha_1 \times \rho$, where $\rho$ is network concentration in total fiber length per volume for the stiff network, and the toughness, which is the area under the stress-strain curve, has the same unit. The data is averaged over five runs and the standard deviations are indicated by errorbars.}
	\label{Fig3}
\end{figure}

To demonstrate, how the fracture mechanics change with the stiff network's proximity to its rigidity percolation threshold, we present results for simulations of the DN close to ($p_1=0.62$) and away from ($p_1=0.80$) the rigidity threshold (Fig.~\ref{Fig3}). The values of $p_1$ were chosen, so that the DN has a finite rigidity, irrespective of $p_2$ \footnote{We found, for example, that when $p_1$ was set to 0.55, the DN had zero shear rigidity and exhibited no stresses when $p_2$ was $0$, $0.2$, or $0.4$}. We find that both the Young's modulus (Fig.\ref{Fig3}(a) and (b)) as well as the stress (Fig. \ref{Fig3}(c) and (d)) developed in the network initially increase with strain and reach a maximum as previously floppy regions become stretched and align to resist deformation. Once the fibers in the network start to experience strains larger than their stretching (or buckling) thresholds, however, they break (or buckle), causing the network to soften. Remarkably, we find that when the stiff network is close to its rigidity threshold, the normalized Young's modulus, the maximum or peak stress, and the strain at failure can be shifted dramatically by the flexible network. This tunability arises because the sparsely populated stiff network allows the DN to undergo non-affine rearrangements \cite{Head:2003,Das:2007,Heussinger:2006}, leading to large variations in rigidity.

We quantify these trends by comparing the peak stress $\sigma_p$, strain at maximum stress $\gamma (\sigma=\sigma_p)$, and the network toughness $\tau$  versus $p_2$ for both DNs in figure~\ref{Fig3}. We find that the peak stress increases with $p_2$ for both DNs due to the additional constraints introduced by the secondary, flexible network. The strain at maximum stress decreases with $p_2$ when the stiff network is close to the percolation threshold and remains nearly constant when the stiff network is far from the percolation threshold. Thus, the additional constraints introduced by the secondary network plays a much greater role in restricting deformation when the stiff network is near the rigidity percolation threshold. Finally, we find that while the network toughness increases for both cases, the increase is greater for the network near the rigidity percolation threshold. This result is somewhat surprising since typically the toughness is proportional to the product of the peak stress and strain at failure which remains nearly constant for the $p_1=0.62$ data. Here, however, because the network fails gradually, the decrease in stress is far less abrupt than in typical materials and the network toughness is substantially increased.    

These results show that the flexible network can modulate the mechanics of the DN far more effectively when the stiff network is just above its rigidity threshold. Additionally, they show how the DN can be modulated to either be extensible, breaking gradually, as is the case for low $p_2$ or be stronger, breaking in a more brittle fashion, as is the case for high $p_2$. The low $p_2$ limit is particularly important in biological tissues such as articular cartilage when it is undergoing osteoarthritis, where the proximity of the stiff (collagen) network to the rigidity percolation threshold varies as a function of tissue depth and the reinforcing flexible network is increasingly removed as the disease progresses\cite{Griffin2014,Jackson2020}. 

\begin{figure}[h]
\includegraphics[width=0.92\columnwidth]{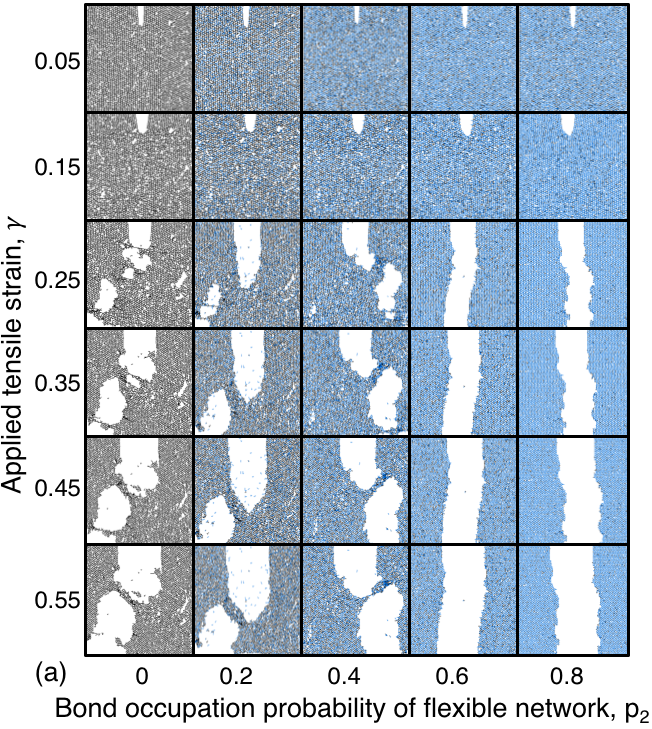}
\includegraphics[width=0.92\columnwidth]{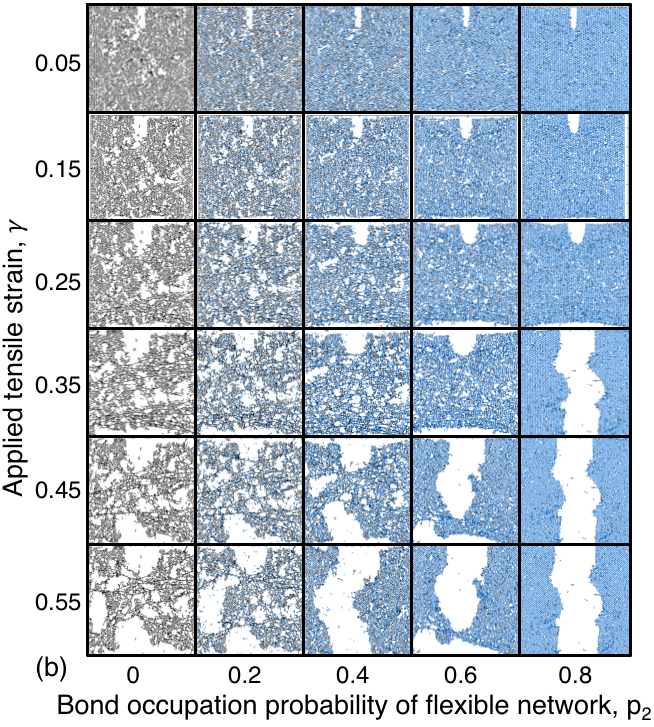}
\caption{Deformation and fracture of the SN and the DN as a function of increasing $p_2$ (x-axis) and applied strain (y-axis). Both the SN and DN were subjected to uniaxial strains applied at the boundary and increased in steps of $1\%$. The value of $p_1$ is set to $0.80$ in (a) and to $0.62$ in (b).}
\label{nets}
\end{figure}

To illustrate the above-mentioned trade-off visually, we present stills from simulations of crack propagation in DNs with $p_1 = 0.62$ and 0.80 as a function of the applied tensile strain $ \gamma $ and $p_2$ (Fig.\ref{nets}). We find that when the stiff network is far above the rigidity threshold ($p_1=0.80$,Fig.\ref{nets}(a)), the DN ruptures abruptly at $\gamma \sim 0.2$ for all $p_2$ values, though the crack morphology is more uniform at higher $p_2$. 
In contrast, when the stiff network is close to the rigidity threshold ($p_1=0.62$,  Fig.\ref{nets}(b)), we observed a wider range of responses. For $p_2 = 0$, 0.2, and 0.4 the networks are extensible, initially developing microcracks that are distributed throughout. With increasing strain these microcracks grow and the network decreases its rigidity while maintaining a percolated structure. For $p_2 = 0.6$ and 0.8 the networks are more brittle, rupturing less homogeneously and maintaining their rigidity up until the point of failure. 

Importantly, this ability to tune the failure characteristics could have numerous important applications. For example, in biologically relevant scenarios the range of tensile strain that a tissue is exposed to may be limited. In this case, it would be more advantageous for the tissue to be extensible rather than brittle over this range of strains. As another example, it may be possible to construct DNs with varying compositions to guide the trajectory or even stall cracks propagating through the material. In fact, given the striking similarity of the crack morphology for $p_2=0.6, \gamma=0.35$ in Fig.~\ref{nets}b to the experimentally observed fracture of articular cartilage tissue (see for example Figure 4 in \cite{Oloyede2003}) it is possible that cartilaginous tissues may already be employing such mechanisms. Finally, we note that the observed richness of behaviors presented here would be further enhanced by including additional tuning parameters such as structural hierarchy, network polarization, or bond polydispersity in either the stiff or flexible networks. This flexibility in resulting material properties and ease of implementation make double networks a very attractive platform for the fabrication of mechanically active artificial tissue constructs. The results presented here are an important step towards achieving this future.     

\begin{acknowledgments}
MD thanks Guy Genin and Markus Linder for useful discussions. This research was supported in part by NSF grants DMR-1808026 and CBET-1604712. 
\end{acknowledgments}

\bibliographystyle{apsrev4-1}
\bibliography{DoubleNetworkBib}

\begin{thebibliography}{37}%
\makeatletter
\providecommand \@ifxundefined [1]{%
 \@ifx{#1\undefined}
}%
\providecommand \@ifnum [1]{%
 \ifnum #1\expandafter \@firstoftwo
 \else \expandafter \@secondoftwo
 \fi
}%
\providecommand \@ifx [1]{%
 \ifx #1\expandafter \@firstoftwo
 \else \expandafter \@secondoftwo
 \fi
}%
\providecommand \natexlab [1]{#1}%
\providecommand \enquote  [1]{``#1''}%
\providecommand \bibnamefont  [1]{#1}%
\providecommand \bibfnamefont [1]{#1}%
\providecommand \citenamefont [1]{#1}%
\providecommand \href@noop [0]{\@secondoftwo}%
\providecommand \href [0]{\begingroup \@sanitize@url \@href}%
\providecommand \@href[1]{\@@startlink{#1}\@@href}%
\providecommand \@@href[1]{\endgroup#1\@@endlink}%
\providecommand \@sanitize@url [0]{\catcode `\\12\catcode `\$12\catcode
  `\&12\catcode `\#12\catcode `\^12\catcode `\_12\catcode `\%12\relax}%
\providecommand \@@startlink[1]{}%
\providecommand \@@endlink[0]{}%
\providecommand \url  [0]{\begingroup\@sanitize@url \@url }%
\providecommand \@url [1]{\endgroup\@href {#1}{\urlprefix }}%
\providecommand \urlprefix  [0]{URL }%
\providecommand \Eprint [0]{\href }%
\providecommand \doibase [0]{http://dx.doi.org/}%
\providecommand \selectlanguage [0]{\@gobble}%
\providecommand \bibinfo  [0]{\@secondoftwo}%
\providecommand \bibfield  [0]{\@secondoftwo}%
\providecommand \translation [1]{[#1]}%
\providecommand \BibitemOpen [0]{}%
\providecommand \bibitemStop [0]{}%
\providecommand \bibitemNoStop [0]{.\EOS\space}%
\providecommand \EOS [0]{\spacefactor3000\relax}%
\providecommand \BibitemShut  [1]{\csname bibitem#1\endcsname}%
\let\auto@bib@innerbib\@empty
\bibitem [{\citenamefont {Pollard}\ and\ \citenamefont
  {D.}(2017)}]{Cytoskeleton}%
  \BibitemOpen
  \bibinfo {editor} {\bibfnamefont {T.~D.}\ \bibnamefont {Pollard}}\ and\
  \bibinfo {editor} {\bibfnamefont {G.~R.}\ \bibnamefont {D.}},\ eds.,\
  \href@noop {} {\emph {\bibinfo {title} {The Cytoskeleton}}}\ (\bibinfo
  {publisher} {Cold Spring Harbor Laboratory Press},\ \bibinfo {year}
  {2017})\BibitemShut {NoStop}%
\bibitem [{\citenamefont {Pelletier}\ \emph {et~al.}(2009)\citenamefont
  {Pelletier}, \citenamefont {Gal}, \citenamefont {Fournier},\ and\
  \citenamefont {Kilfoil}}]{Pelletier}%
  \BibitemOpen
  \bibfield  {author} {\bibinfo {author} {\bibfnamefont {V.}~\bibnamefont
  {Pelletier}}, \bibinfo {author} {\bibfnamefont {N.}~\bibnamefont {Gal}},
  \bibinfo {author} {\bibfnamefont {P.}~\bibnamefont {Fournier}}, \ and\
  \bibinfo {author} {\bibfnamefont {M.~L.}\ \bibnamefont {Kilfoil}},\ }\href
  {\doibase 10.1103/PhysRevLett.102.188303} {\bibfield  {journal} {\bibinfo
  {journal} {Phys. Rev. Lett.}\ }\textbf {\bibinfo {volume} {102}},\ \bibinfo
  {pages} {188303} (\bibinfo {year} {2009})}\BibitemShut {NoStop}%
\bibitem [{\citenamefont {Shea~N.Ricketts}\ and\ \citenamefont
  {M.Robertson-Anderson}(2018)}]{Ricketts2018}%
  \BibitemOpen
  \bibfield  {author} {\bibinfo {author} {\bibfnamefont {J.~L.~R.}\
  \bibnamefont {Shea~N.Ricketts}}\ and\ \bibinfo {author} {\bibfnamefont
  {R.}~\bibnamefont {M.Robertson-Anderson}},\ }\href {\doibase
  https://doi.org/10.1016/j.bpj.2018.08.010} {\bibfield  {journal} {\bibinfo
  {journal} {Biophysical Journal}\ }\textbf {\bibinfo {volume} {115}},\
  \bibinfo {pages} {1055 } (\bibinfo {year} {2018})}\BibitemShut {NoStop}%
\bibitem [{\citenamefont {Ricketts}\ \emph {et~al.}(2019)\citenamefont
  {Ricketts}, \citenamefont {Francis}, \citenamefont {Farhadi}, \citenamefont
  {Rust}, \citenamefont {Das}, \citenamefont {Ross},\ and\ \citenamefont
  {Robertson-Anderson}}]{Ricketts2019}%
  \BibitemOpen
  \bibfield  {author} {\bibinfo {author} {\bibfnamefont {S.~N.}\ \bibnamefont
  {Ricketts}}, \bibinfo {author} {\bibfnamefont {M.~L.}\ \bibnamefont
  {Francis}}, \bibinfo {author} {\bibfnamefont {L.}~\bibnamefont {Farhadi}},
  \bibinfo {author} {\bibfnamefont {M.~J.}\ \bibnamefont {Rust}}, \bibinfo
  {author} {\bibfnamefont {M.}~\bibnamefont {Das}}, \bibinfo {author}
  {\bibfnamefont {J.~L.}\ \bibnamefont {Ross}}, \ and\ \bibinfo {author}
  {\bibfnamefont {R.~M.}\ \bibnamefont {Robertson-Anderson}},\ }\href {\doibase
  10.1038/s41598-019-49236-4} {\bibfield  {journal} {\bibinfo  {journal}
  {Scientific Reports}\ }\textbf {\bibinfo {volume} {9}} (\bibinfo {year}
  {2019}),\ 10.1038/s41598-019-49236-4}\BibitemShut {NoStop}%
\bibitem [{\citenamefont {Farhadi}\ \emph {et~al.}(2020)\citenamefont
  {Farhadi}, \citenamefont {Ricketts}, \citenamefont {Rust}, \citenamefont
  {Das}, \citenamefont {Robertson-Anderson},\ and\ \citenamefont
  {Ross}}]{Farhadi}%
  \BibitemOpen
  \bibfield  {author} {\bibinfo {author} {\bibfnamefont {L.}~\bibnamefont
  {Farhadi}}, \bibinfo {author} {\bibfnamefont {S.~N.}\ \bibnamefont
  {Ricketts}}, \bibinfo {author} {\bibfnamefont {M.~J.}\ \bibnamefont {Rust}},
  \bibinfo {author} {\bibfnamefont {M.}~\bibnamefont {Das}}, \bibinfo {author}
  {\bibfnamefont {R.~M.}\ \bibnamefont {Robertson-Anderson}}, \ and\ \bibinfo
  {author} {\bibfnamefont {J.~L.}\ \bibnamefont {Ross}},\ }\href {\doibase
  10.1039/C9SM02400J} {\bibfield  {journal} {\bibinfo  {journal} {Soft Matter}\
  ,\ } (\bibinfo {year} {2020})}\BibitemShut {NoStop}%
\bibitem [{\citenamefont {Mow}\ \emph {et~al.}(1992)\citenamefont {Mow},
  \citenamefont {Ratcliffe},\ and\ \citenamefont {Poole}}]{Mow:1992}%
  \BibitemOpen
  \bibfield  {author} {\bibinfo {author} {\bibfnamefont {V.~C.}\ \bibnamefont
  {Mow}}, \bibinfo {author} {\bibfnamefont {A.}~\bibnamefont {Ratcliffe}}, \
  and\ \bibinfo {author} {\bibfnamefont {A.~R.}\ \bibnamefont {Poole}},\ }\href
  {\doibase 10.1016/0142-9612(92)90001-5} {\bibfield  {journal} {\bibinfo
  {journal} {{B}iomaterials}\ }\textbf {\bibinfo {volume} {13}},\ \bibinfo
  {pages} {67} (\bibinfo {year} {1992})}\BibitemShut {NoStop}%
\bibitem [{\citenamefont {Fung}(1993)}]{Fung:1993}%
  \BibitemOpen
  \bibfield  {author} {\bibinfo {author} {\bibfnamefont {Y.}~\bibnamefont
  {Fung}},\ }\href@noop {} {\emph {\bibinfo {title} {Biomechanics: Mechanical
  Properties of Living Tissues}}},\ \bibinfo {edition} {2nd}\ ed.\ (\bibinfo
  {publisher} {Springer-Verlag},\ \bibinfo {address} {New York},\ \bibinfo
  {year} {1993})\BibitemShut {NoStop}%
\bibitem [{\citenamefont {O'Driscoll}(1998)}]{ODriscoll:1998}%
  \BibitemOpen
  \bibfield  {author} {\bibinfo {author} {\bibfnamefont {S.~W.}\ \bibnamefont
  {O'Driscoll}},\ }\href@noop {} {\bibfield  {journal} {\bibinfo  {journal}
  {{J}. {B}one {J}oint {S}urg. {A}m.}\ }\textbf {\bibinfo {volume} {80}},\
  \bibinfo {pages} {1795} (\bibinfo {year} {1998})}\BibitemShut {NoStop}%
\bibitem [{\citenamefont {Lai}\ and\ \citenamefont
  {Levenston}(2010)}]{Lai:2010}%
  \BibitemOpen
  \bibfield  {author} {\bibinfo {author} {\bibfnamefont {J.~H.}\ \bibnamefont
  {Lai}}\ and\ \bibinfo {author} {\bibfnamefont {M.~E.}\ \bibnamefont
  {Levenston}},\ }\href@noop {} {\bibfield  {journal} {\bibinfo  {journal}
  {Osteoarthritis Cartilage}\ }\textbf {\bibinfo {volume} {18}},\ \bibinfo
  {pages} {1291} (\bibinfo {year} {2010})}\BibitemShut {NoStop}%
\bibitem [{\citenamefont {Haque}\ \emph {et~al.}(2012)\citenamefont {Haque},
  \citenamefont {Kurokawa},\ and\ \citenamefont {Gong}}]{Gong2012}%
  \BibitemOpen
  \bibfield  {author} {\bibinfo {author} {\bibfnamefont {M.~A.}\ \bibnamefont
  {Haque}}, \bibinfo {author} {\bibfnamefont {T.}~\bibnamefont {Kurokawa}}, \
  and\ \bibinfo {author} {\bibfnamefont {J.~P.}\ \bibnamefont {Gong}},\
  }\href@noop {} {\bibfield  {journal} {\bibinfo  {journal} {Polymer}\ }\textbf
  {\bibinfo {volume} {53}},\ \bibinfo {pages} {1805} (\bibinfo {year}
  {2012})}\BibitemShut {NoStop}%
\bibitem [{\citenamefont {Nonoyama}\ and\ \citenamefont
  {Gong}(2015)}]{Gong2015}%
  \BibitemOpen
  \bibfield  {author} {\bibinfo {author} {\bibfnamefont {T.}~\bibnamefont
  {Nonoyama}}\ and\ \bibinfo {author} {\bibfnamefont {J.~P.}\ \bibnamefont
  {Gong}},\ }\href@noop {} {\bibfield  {journal} {\bibinfo  {journal}
  {Polymer}\ }\textbf {\bibinfo {volume} {229}},\ \bibinfo {pages} {853}
  (\bibinfo {year} {2015})}\BibitemShut {NoStop}%
\bibitem [{\citenamefont {Broedersz}\ \emph {et~al.}(2011)\citenamefont
  {Broedersz}, \citenamefont {Mao}, \citenamefont {Lubensky},\ and\
  \citenamefont {MacKintosh}}]{Broedersz:2011}%
  \BibitemOpen
  \bibfield  {author} {\bibinfo {author} {\bibfnamefont {C.}~\bibnamefont
  {Broedersz}}, \bibinfo {author} {\bibfnamefont {X.}~\bibnamefont {Mao}},
  \bibinfo {author} {\bibfnamefont {T.}~\bibnamefont {Lubensky}}, \ and\
  \bibinfo {author} {\bibfnamefont {F.~C.}\ \bibnamefont {MacKintosh}},\ }\href
  {\doibase 10.1038/nphys2127} {\bibfield  {journal} {\bibinfo  {journal}
  {Nature Phys.}\ }\textbf {\bibinfo {volume} {7}},\ \bibinfo {pages} {983}
  (\bibinfo {year} {2011})}\BibitemShut {NoStop}%
\bibitem [{\citenamefont {Head}\ \emph {et~al.}(2003)\citenamefont {Head},
  \citenamefont {Levine},\ and\ \citenamefont {MacKintosh}}]{Head:2003}%
  \BibitemOpen
  \bibfield  {author} {\bibinfo {author} {\bibfnamefont {D.~A.}\ \bibnamefont
  {Head}}, \bibinfo {author} {\bibfnamefont {A.~J.}\ \bibnamefont {Levine}}, \
  and\ \bibinfo {author} {\bibfnamefont {F.~C.}\ \bibnamefont {MacKintosh}},\
  }\href {\doibase 10.1103/PhysRevE.68.061907} {\bibfield  {journal} {\bibinfo
  {journal} {{P}hys. {R}ev. E}\ }\textbf {\bibinfo {volume} {68}},\ \bibinfo
  {pages} {061907} (\bibinfo {year} {2003})}\BibitemShut {NoStop}%
\bibitem [{\citenamefont {Wilhelm}\ and\ \citenamefont
  {Frey}(2003)}]{Wilhelm:2003}%
  \BibitemOpen
  \bibfield  {author} {\bibinfo {author} {\bibfnamefont {J.}~\bibnamefont
  {Wilhelm}}\ and\ \bibinfo {author} {\bibfnamefont {E.}~\bibnamefont {Frey}},\
  }\href {\doibase 10.1103/PhysRevE.68.061907} {\bibfield  {journal} {\bibinfo
  {journal} {Phys. Rev. Lett.}\ }\textbf {\bibinfo {volume} {68}},\ \bibinfo
  {pages} {061907} (\bibinfo {year} {2003})}\BibitemShut {NoStop}%
\bibitem [{\citenamefont {Gardel}\ \emph {et~al.}(2004)\citenamefont {Gardel},
  \citenamefont {Shin}, \citenamefont {MacKintosh}, \citenamefont {Mahadevan},
  \citenamefont {Matsudaira},\ and\ \citenamefont {Weitz}}]{gardel2004scaling}%
  \BibitemOpen
  \bibfield  {author} {\bibinfo {author} {\bibfnamefont {M.}~\bibnamefont
  {Gardel}}, \bibinfo {author} {\bibfnamefont {J.}~\bibnamefont {Shin}},
  \bibinfo {author} {\bibfnamefont {F.}~\bibnamefont {MacKintosh}}, \bibinfo
  {author} {\bibfnamefont {L.}~\bibnamefont {Mahadevan}}, \bibinfo {author}
  {\bibfnamefont {P.}~\bibnamefont {Matsudaira}}, \ and\ \bibinfo {author}
  {\bibfnamefont {D.}~\bibnamefont {Weitz}},\ }\href@noop {} {\bibfield
  {journal} {\bibinfo  {journal} {Physical review letters}\ }\textbf {\bibinfo
  {volume} {93}},\ \bibinfo {pages} {188102} (\bibinfo {year}
  {2004})}\BibitemShut {NoStop}%
\bibitem [{\citenamefont {Missel}\ \emph {et~al.}(2010)\citenamefont {Missel},
  \citenamefont {Bai}, \citenamefont {Klug},\ and\ \citenamefont
  {Levine}}]{Missel:2010}%
  \BibitemOpen
  \bibfield  {author} {\bibinfo {author} {\bibfnamefont {A.~R.}\ \bibnamefont
  {Missel}}, \bibinfo {author} {\bibfnamefont {M.}~\bibnamefont {Bai}},
  \bibinfo {author} {\bibfnamefont {W.~S.}\ \bibnamefont {Klug}}, \ and\
  \bibinfo {author} {\bibfnamefont {A.~J.}\ \bibnamefont {Levine}},\
  }\href@noop {} {\bibfield  {journal} {\bibinfo  {journal} {Phys. Rev. E}\
  }\textbf {\bibinfo {volume} {82}},\ \bibinfo {pages} {041907} (\bibinfo
  {year} {2010})}\BibitemShut {NoStop}%
\bibitem [{\citenamefont {Heussinger}\ and\ \citenamefont
  {Frey}(2006)}]{Heussinger:2006}%
  \BibitemOpen
  \bibfield  {author} {\bibinfo {author} {\bibfnamefont {C.}~\bibnamefont
  {Heussinger}}\ and\ \bibinfo {author} {\bibfnamefont {E.}~\bibnamefont
  {Frey}},\ }\href@noop {} {\bibfield  {journal} {\bibinfo  {journal} {Phys.
  Rev. Lett.}\ }\textbf {\bibinfo {volume} {97}},\ \bibinfo {pages} {105501}
  (\bibinfo {year} {2006})}\BibitemShut {NoStop}%
\bibitem [{\citenamefont {Das}\ \emph {et~al.}(2007)\citenamefont {Das},
  \citenamefont {Mackintosh},\ and\ \citenamefont {Levine}}]{Das:2007}%
  \BibitemOpen
  \bibfield  {author} {\bibinfo {author} {\bibfnamefont {M.}~\bibnamefont
  {Das}}, \bibinfo {author} {\bibfnamefont {F.~C.}\ \bibnamefont {Mackintosh}},
  \ and\ \bibinfo {author} {\bibfnamefont {A.~J.}\ \bibnamefont {Levine}},\
  }\href@noop {} {\bibfield  {journal} {\bibinfo  {journal} {Phys. Rev. Lett.}\
  }\textbf {\bibinfo {volume} {99}},\ \bibinfo {pages} {038101} (\bibinfo
  {year} {2007})}\BibitemShut {NoStop}%
\bibitem [{\citenamefont {Kang}\ \emph {et~al.}(2009)\citenamefont {Kang},
  \citenamefont {Wen}, \citenamefont {Janmey}, \citenamefont {Tang},
  \citenamefont {Conti},\ and\ \citenamefont {MacKintosh}}]{kang2009nonlinear}%
  \BibitemOpen
  \bibfield  {author} {\bibinfo {author} {\bibfnamefont {H.}~\bibnamefont
  {Kang}}, \bibinfo {author} {\bibfnamefont {Q.}~\bibnamefont {Wen}}, \bibinfo
  {author} {\bibfnamefont {P.~A.}\ \bibnamefont {Janmey}}, \bibinfo {author}
  {\bibfnamefont {J.~X.}\ \bibnamefont {Tang}}, \bibinfo {author}
  {\bibfnamefont {E.}~\bibnamefont {Conti}}, \ and\ \bibinfo {author}
  {\bibfnamefont {F.~C.}\ \bibnamefont {MacKintosh}},\ }\href@noop {}
  {\bibfield  {journal} {\bibinfo  {journal} {The Journal of Physical Chemistry
  B}\ }\textbf {\bibinfo {volume} {113}},\ \bibinfo {pages} {3799} (\bibinfo
  {year} {2009})}\BibitemShut {NoStop}%
\bibitem [{\citenamefont {Sheinman}\ \emph {et~al.}(2012)\citenamefont
  {Sheinman}, \citenamefont {Broedersz},\ and\ \citenamefont
  {MacKintosh}}]{Sheinman:2012}%
  \BibitemOpen
  \bibfield  {author} {\bibinfo {author} {\bibfnamefont {M.}~\bibnamefont
  {Sheinman}}, \bibinfo {author} {\bibfnamefont {C.~P.}\ \bibnamefont
  {Broedersz}}, \ and\ \bibinfo {author} {\bibfnamefont {F.~C.}\ \bibnamefont
  {MacKintosh}},\ }\href@noop {} {\bibfield  {journal} {\bibinfo  {journal}
  {Phys. Rev. E}\ }\textbf {\bibinfo {volume} {85}},\ \bibinfo {pages} {021801}
  (\bibinfo {year} {2012})}\BibitemShut {NoStop}%
\bibitem [{\citenamefont {Picu}(2011)}]{Picu:2011}%
  \BibitemOpen
  \bibfield  {author} {\bibinfo {author} {\bibfnamefont {R.~C.}\ \bibnamefont
  {Picu}},\ }\href {\doibase 10.1039/C1SM05022B} {\bibfield  {journal}
  {\bibinfo  {journal} {Soft Matter}\ }\textbf {\bibinfo {volume} {7}},\
  \bibinfo {pages} {6768} (\bibinfo {year} {2011})}\BibitemShut {NoStop}%
\bibitem [{\citenamefont {Zhang}\ \emph {et~al.}(2017)\citenamefont {Zhang},
  \citenamefont {Rocklin}, \citenamefont {Sander},\ and\ \citenamefont
  {Mao}}]{Zhang2019}%
  \BibitemOpen
  \bibfield  {author} {\bibinfo {author} {\bibfnamefont {L.}~\bibnamefont
  {Zhang}}, \bibinfo {author} {\bibfnamefont {D.~Z.}\ \bibnamefont {Rocklin}},
  \bibinfo {author} {\bibfnamefont {L.~M.}\ \bibnamefont {Sander}}, \ and\
  \bibinfo {author} {\bibfnamefont {X.}~\bibnamefont {Mao}},\ }\href {\doibase
  10.1103/PhysRevMaterials.1.052602} {\bibfield  {journal} {\bibinfo  {journal}
  {Phys. Rev. Materials}\ }\textbf {\bibinfo {volume} {1}},\ \bibinfo {pages}
  {052602} (\bibinfo {year} {2017})}\BibitemShut {NoStop}%
\bibitem [{\citenamefont {Burla}\ \emph {et~al.}(2020)\citenamefont {Burla},
  \citenamefont {Dussi}, \citenamefont {Martinez-Torres}, \citenamefont
  {Tauber}, \citenamefont {van~der Gucht},\ and\ \citenamefont
  {Koenderink}}]{Burla2020}%
  \BibitemOpen
  \bibfield  {author} {\bibinfo {author} {\bibfnamefont {F.}~\bibnamefont
  {Burla}}, \bibinfo {author} {\bibfnamefont {S.}~\bibnamefont {Dussi}},
  \bibinfo {author} {\bibfnamefont {C.}~\bibnamefont {Martinez-Torres}},
  \bibinfo {author} {\bibfnamefont {J.}~\bibnamefont {Tauber}}, \bibinfo
  {author} {\bibfnamefont {J.}~\bibnamefont {van~der Gucht}}, \ and\ \bibinfo
  {author} {\bibfnamefont {G.~H.}\ \bibnamefont {Koenderink}},\ }\href
  {\doibase 10.1073/pnas.1920062117} {\bibfield  {journal} {\bibinfo  {journal}
  {Proceedings of the National Academy of Sciences}\ }\textbf {\bibinfo
  {volume} {117}},\ \bibinfo {pages} {8326} (\bibinfo {year}
  {2020})}\BibitemShut {NoStop}%
\bibitem [{\citenamefont {Das}\ \emph {et~al.}(2008)\citenamefont {Das},
  \citenamefont {Levine},\ and\ \citenamefont {Mackintosh}}]{Das:2008}%
  \BibitemOpen
  \bibfield  {author} {\bibinfo {author} {\bibfnamefont {M.}~\bibnamefont
  {Das}}, \bibinfo {author} {\bibfnamefont {A.~J.}\ \bibnamefont {Levine}}, \
  and\ \bibinfo {author} {\bibfnamefont {F.~C.}\ \bibnamefont {Mackintosh}},\
  }\href@noop {} {\bibfield  {journal} {\bibinfo  {journal} {Europhys. Lett.}\
  }\textbf {\bibinfo {volume} {84}},\ \bibinfo {pages} {18003} (\bibinfo {year}
  {2008})}\BibitemShut {NoStop}%
\bibitem [{\citenamefont {Das}\ and\ \citenamefont
  {MacKintosh}(2010)}]{Das:2010}%
  \BibitemOpen
  \bibfield  {author} {\bibinfo {author} {\bibfnamefont {M.}~\bibnamefont
  {Das}}\ and\ \bibinfo {author} {\bibfnamefont {F.~C.}\ \bibnamefont
  {MacKintosh}},\ }\href {\doibase 10.1103/PhysRevLett.105.138102} {\bibfield
  {journal} {\bibinfo  {journal} {Phys. Rev. Lett.}\ }\textbf {\bibinfo
  {volume} {105}},\ \bibinfo {pages} {138102} (\bibinfo {year}
  {2010})}\BibitemShut {NoStop}%
\bibitem [{\citenamefont {Burla}\ \emph {et~al.}(2019)\citenamefont {Burla},
  \citenamefont {Tauber}, \citenamefont {Dussi}, \citenamefont {van~der
  Gucht},\ and\ \citenamefont {Koenderink}}]{Burla2019}%
  \BibitemOpen
  \bibfield  {author} {\bibinfo {author} {\bibfnamefont {F.}~\bibnamefont
  {Burla}}, \bibinfo {author} {\bibfnamefont {J.}~\bibnamefont {Tauber}},
  \bibinfo {author} {\bibfnamefont {S.}~\bibnamefont {Dussi}}, \bibinfo
  {author} {\bibfnamefont {J.}~\bibnamefont {van~der Gucht}}, \ and\ \bibinfo
  {author} {\bibfnamefont {G.~H.}\ \bibnamefont {Koenderink}},\ }\href
  {\doibase 10.1038/s41567-019-0443-6} {\bibfield  {journal} {\bibinfo
  {journal} {Nature Physics}\ }\textbf {\bibinfo {volume} {15}},\ \bibinfo
  {pages} {549} (\bibinfo {year} {2019})}\BibitemShut {NoStop}%
\bibitem [{\citenamefont {Trivedi}\ \emph {et~al.}(2008)\citenamefont
  {Trivedi}, \citenamefont {Rahn}, \citenamefont {Kier},\ and\ \citenamefont
  {Walker}}]{Trivedia:2008}%
  \BibitemOpen
  \bibfield  {author} {\bibinfo {author} {\bibfnamefont {D.}~\bibnamefont
  {Trivedi}}, \bibinfo {author} {\bibfnamefont {C.~D.}\ \bibnamefont {Rahn}},
  \bibinfo {author} {\bibfnamefont {W.~M.}\ \bibnamefont {Kier}}, \ and\
  \bibinfo {author} {\bibfnamefont {I.~D.}\ \bibnamefont {Walker}},\ }\href
  {\doibase 10.1080/11762320802557865} {\bibfield  {journal} {\bibinfo
  {journal} {{A}ppl. {B}ion. {B}iomech.}\ }\textbf {\bibinfo {volume} {5}},\
  \bibinfo {pages} {99} (\bibinfo {year} {2008})}\BibitemShut {NoStop}%
\bibitem [{\citenamefont {Taylor}\ \emph {et~al.}(2012)\citenamefont {Taylor},
  \citenamefont {O'Mara}, \citenamefont {Ryan}, \citenamefont {Takaza},\ and\
  \citenamefont {Simms}}]{Simms}%
  \BibitemOpen
  \bibfield  {author} {\bibinfo {author} {\bibfnamefont {D.}~\bibnamefont
  {Taylor}}, \bibinfo {author} {\bibfnamefont {N.}~\bibnamefont {O'Mara}},
  \bibinfo {author} {\bibfnamefont {E.}~\bibnamefont {Ryan}}, \bibinfo {author}
  {\bibfnamefont {M.}~\bibnamefont {Takaza}}, \ and\ \bibinfo {author}
  {\bibfnamefont {C.}~\bibnamefont {Simms}},\ }\href@noop {} {\bibfield
  {journal} {\bibinfo  {journal} {Journal of the mechanical behavior of
  biomedical materials}\ }\textbf {\bibinfo {volume} {6}},\ \bibinfo {pages}
  {139} (\bibinfo {year} {2012})}\BibitemShut {NoStop}%
\bibitem [{\citenamefont {Silverberg}\ \emph {et~al.}(2014)\citenamefont
  {Silverberg}, \citenamefont {Barrett}, \citenamefont {Das}, \citenamefont
  {Petersen}, \citenamefont {Bonassar},\ and\ \citenamefont
  {Cohen}}]{silverberg2014structure}%
  \BibitemOpen
  \bibfield  {author} {\bibinfo {author} {\bibfnamefont {J.~L.}\ \bibnamefont
  {Silverberg}}, \bibinfo {author} {\bibfnamefont {A.~R.}\ \bibnamefont
  {Barrett}}, \bibinfo {author} {\bibfnamefont {M.}~\bibnamefont {Das}},
  \bibinfo {author} {\bibfnamefont {P.~B.}\ \bibnamefont {Petersen}}, \bibinfo
  {author} {\bibfnamefont {L.~J.}\ \bibnamefont {Bonassar}}, \ and\ \bibinfo
  {author} {\bibfnamefont {I.}~\bibnamefont {Cohen}},\ }\href@noop {}
  {\bibfield  {journal} {\bibinfo  {journal} {Biophysical journal}\ }\textbf
  {\bibinfo {volume} {107}},\ \bibinfo {pages} {1721} (\bibinfo {year}
  {2014})}\BibitemShut {NoStop}%
\bibitem [{\citenamefont {Feng}\ \emph {et~al.}(1985)\citenamefont {Feng},
  \citenamefont {Thorpe},\ and\ \citenamefont {Garboczi}}]{FengThorpe}%
  \BibitemOpen
  \bibfield  {author} {\bibinfo {author} {\bibfnamefont {S.}~\bibnamefont
  {Feng}}, \bibinfo {author} {\bibfnamefont {M.~F.}\ \bibnamefont {Thorpe}}, \
  and\ \bibinfo {author} {\bibfnamefont {E.}~\bibnamefont {Garboczi}},\ }\href
  {\doibase 10.1103/PhysRevB.31.276} {\bibfield  {journal} {\bibinfo  {journal}
  {Phys. Rev. B}\ }\textbf {\bibinfo {volume} {31}},\ \bibinfo {pages} {276}
  (\bibinfo {year} {1985})}\BibitemShut {NoStop}%
\bibitem [{\citenamefont {Das}\ \emph {et~al.}(2012)\citenamefont {Das},
  \citenamefont {Quint},\ and\ \citenamefont {Schwarz}}]{Das:2012}%
  \BibitemOpen
  \bibfield  {author} {\bibinfo {author} {\bibfnamefont {M.}~\bibnamefont
  {Das}}, \bibinfo {author} {\bibfnamefont {D.~A.}\ \bibnamefont {Quint}}, \
  and\ \bibinfo {author} {\bibfnamefont {J.~M.}\ \bibnamefont {Schwarz}},\
  }\href {\doibase 10.1371/journal.pone.0035939} {\bibfield  {journal}
  {\bibinfo  {journal} {{PL}o{S} {ONE}}\ }\textbf {\bibinfo {volume} {7}},\
  \bibinfo {pages} {e35939} (\bibinfo {year} {2012})}\BibitemShut {NoStop}%
\bibitem [{Note1()}]{Note1}%
  \BibitemOpen
  \bibinfo {note} {The value of $\alpha _3$ was chosen to be the effective
  spring constant of two springs $\alpha _1$ and $\alpha _2$ in parallel.
  Simulations with a much small value of $\alpha _3/\alpha _1=0.01$ yielded
  qualitatively similar results and the percolation thresholds were
  unchanged.}\BibitemShut {Stop}%
\bibitem [{\citenamefont {Mao}\ \emph {et~al.}(2013)\citenamefont {Mao},
  \citenamefont {Stenull},\ and\ \citenamefont {Lubensky}}]{Lubensky2013}%
  \BibitemOpen
  \bibfield  {author} {\bibinfo {author} {\bibfnamefont {X.}~\bibnamefont
  {Mao}}, \bibinfo {author} {\bibfnamefont {O.}~\bibnamefont {Stenull}}, \ and\
  \bibinfo {author} {\bibfnamefont {T.~C.}\ \bibnamefont {Lubensky}},\ }\href
  {\doibase 10.1103/PhysRevE.87.042602} {\bibfield  {journal} {\bibinfo
  {journal} {Phys. Rev. E}\ }\textbf {\bibinfo {volume} {87}},\ \bibinfo
  {pages} {042602} (\bibinfo {year} {2013})}\BibitemShut {NoStop}%
\bibitem [{\citenamefont {Stok}\ and\ \citenamefont
  {Oloyede}(2003)}]{Oloyede2003}%
  \BibitemOpen
  \bibfield  {author} {\bibinfo {author} {\bibfnamefont {K.}~\bibnamefont
  {Stok}}\ and\ \bibinfo {author} {\bibfnamefont {A.}~\bibnamefont {Oloyede}},\
  }\href@noop {} {\bibfield  {journal} {\bibinfo  {journal} {Connective Tissue
  Research}\ }\textbf {\bibinfo {volume} {44}},\ \bibinfo {pages} {109}
  (\bibinfo {year} {2003})}\BibitemShut {NoStop}%
\bibitem [{Note2()}]{Note2}%
  \BibitemOpen
  \bibinfo {note} {We found, for example, that when $p_1$ was set to 0.55, the
  DN had zero shear rigidity and exhibited no stresses when $p_2$ was $0$,
  $0.2$, or $0.4$}\BibitemShut {NoStop}%
\bibitem [{\citenamefont {Griffin}\ \emph {et~al.}(2014)\citenamefont
  {Griffin}, \citenamefont {Vicari}, \citenamefont {Buckley}, \citenamefont
  {Silverberg}, \citenamefont {Cohen},\ and\ \citenamefont
  {Bonassar}}]{Griffin2014}%
  \BibitemOpen
  \bibfield  {author} {\bibinfo {author} {\bibfnamefont {D.~J.}\ \bibnamefont
  {Griffin}}, \bibinfo {author} {\bibfnamefont {J.}~\bibnamefont {Vicari}},
  \bibinfo {author} {\bibfnamefont {M.~R.}\ \bibnamefont {Buckley}}, \bibinfo
  {author} {\bibfnamefont {J.~L.}\ \bibnamefont {Silverberg}}, \bibinfo
  {author} {\bibfnamefont {I.}~\bibnamefont {Cohen}}, \ and\ \bibinfo {author}
  {\bibfnamefont {L.~J.}\ \bibnamefont {Bonassar}},\ }\href {\doibase
  10.1002/jor.22713} {\bibfield  {journal} {\bibinfo  {journal} {Journal of
  Orthopaedic Research}\ }\textbf {\bibinfo {volume} {32}},\ \bibinfo {pages}
  {1652} (\bibinfo {year} {2014})},\ \Eprint
  {http://arxiv.org/abs/https://onlinelibrary.wiley.com/doi/pdf/10.1002/jor.22713}
  {https://onlinelibrary.wiley.com/doi/pdf/10.1002/jor.22713} \BibitemShut
  {NoStop}%
\bibitem [{\citenamefont {Jackson}\ \emph {et~al.}()\citenamefont {Jackson},
  \citenamefont {Das}, \citenamefont {Bartell}, \citenamefont {Fourtier},
  \citenamefont {Bonassar},\ and\ \citenamefont {Cohen}}]{Jackson2020}%
  \BibitemOpen
  \bibfield  {author} {\bibinfo {author} {\bibfnamefont {T.~W.}\ \bibnamefont
  {Jackson}}, \bibinfo {author} {\bibfnamefont {M.}~\bibnamefont {Das}},
  \bibinfo {author} {\bibfnamefont {L.}~\bibnamefont {Bartell}}, \bibinfo
  {author} {\bibfnamefont {L.}~\bibnamefont {Fourtier}}, \bibinfo {author}
  {\bibfnamefont {L.~J.}\ \bibnamefont {Bonassar}}, \ and\ \bibinfo {author}
  {\bibfnamefont {I.}~\bibnamefont {Cohen}},\ }\href@noop {} {\enquote
  {\bibinfo {title} {Enhanced sensitivity of cartilage shear mechanics to
  aggrecan concentration near the collagen rigidity percolation threshold},}\
  }\bibinfo {note} {Unpublished}\BibitemShut {NoStop}%
\end{thebibliography}%

\end{document}